\documentclass[prl,twocolumn,letterpaper,superscriptaddress,reprint,aps]{revtex4-1}

\usepackage{threeparttable}
\usepackage{graphicx}
\usepackage{pslatex}
\usepackage{amsmath}
\usepackage{booktabs, caption, makecell}
\newcommand{\atUCLA}{Dept. of Physics and Astronomy, Univ. of California, Los Angeles, Los Angeles, CA 90095.}
\newcommand{\atOSU}{Dept. of Physics, Center for Cosmology and AstroParticle Physics, Ohio State Univ., Columbus, OH 43210.}
\newcommand{\atUH}{Dept. of Physics and Astronomy, Univ. of Hawaii, Manoa, HI 96822.}
\newcommand{\atNTU}{Dept. of Physics, Grad. Inst. of Astrophys.,\& Leung Center for Cosmology and Particle Astrophysics, National Taiwan University, Taipei, Taiwan.}
\newcommand{\atTungU}{Dept. of Electrophysics, National Yang-Ming Chiao Tung University, Hsinchu 30010, Taiwan.}
\newcommand{\atKU}{Dept. of Physics and Astronomy, Univ. of Kansas, Lawrence, KS 66045.}
\newcommand{\atWashU}{Dept of Physics \& McDonnell Center for the Space Sciences, Washington Univ in St Louis, MO, 63130}
\newcommand{\atSLAC}{SLAC National Accelerator Laboratory, Menlo Park, CA, 94025.}
\newcommand{\atUD}{Dept. of Physics, Univ. of Delaware, Newark, DE 19716.}
\newcommand{\atUCL}{Dept. of Physics and Astronomy, University College London, London, United Kingdom.}

\newcommand{\atJPL}{Jet Propulsion Laboratory, California Institute of Technology, Pasadena, CA 91109.}

\newcommand{\atChicago}{Dept. of Physics, Enrico Fermi Institute, Kavli Institute for Cosmological Physics, Univ. of Chicago , Chicago IL 60637.}

\newcommand{\atPennState}{Dept. of Physics, Dept. of Astronomy \& Astrophysics, Penn State Univ., University Park, PA 16801}
\newcommand{\atGWU}{Dept. of Mathematics, George Washington University, Washington D.C.}
\newcommand{\atMPI}{Max-Planck-Institute f\:ur Kernphysik: Heidelberg, Germany.}

\begin{document}

\title{Unusual Near-horizon Cosmic-ray-like Events Observed by ANITA-IV}

\author{P.~W.~Gorham}
\affiliation{\atUH}

\author{A.~Ludwig}
\affiliation{\atChicago}

\author{C.~Deaconu}
\affiliation{\atChicago}

\author{P.~Cao}
\affiliation{\atUD}

\author{P.~Allison}
\affiliation{\atOSU}

\author{O.~Banerjee}
\affiliation{\atOSU}

\author{L.~Batten}
\affiliation{\atUCL}

\author{D.~Bhattacharya}
\affiliation{\atGWU}

\author{J.~J.~Beatty}
\affiliation{\atOSU}

\author{K.~Belov}
\affiliation{\atJPL}


\author{W.~R.~Binns}
\affiliation{\atWashU}

\author{V.~Bugaev}
\affiliation{\atWashU}

\author{C.~H.~Chen}
\affiliation{\atNTU}

\author{P.~Chen}
\affiliation{\atNTU}

\author{Y.~Chen}
\affiliation{\atNTU}

\author{J.~M.~Clem}
\affiliation{\atUD}
%

\author{L.~Cremonesi}
\affiliation{\atUCL}

\author{B.~Dailey}
\affiliation{\atOSU}

\author{P.~F.~Dowkontt}
\affiliation{\atWashU}

\author{B.~D.~Fox}
\affiliation{\atUH}

\author{J.~W.~H.~Gordon}
\affiliation{\atOSU}

\author{C.~Hast}
\affiliation{\atSLAC}

\author{B.~Hill}
\affiliation{\atUH}

\author{S.~Y. Hsu}
\affiliation{\atNTU}

\author{J.~J.~Huang}
\affiliation{\atNTU}

\author{K.~Hughes}
\affiliation{\atOSU}
\altaffiliation{\atChicago}

\author{R.~Hupe}
\affiliation{\atOSU}

\author{M.~H.~Israel}
\affiliation{\atWashU}

\author{T.C. Liu}
\affiliation{\atTungU}

\author{L.~Macchiarulo}
\affiliation{\atUH}

\author{S.~Matsuno}
\affiliation{\atUH}

\author{K.~McBride}
\affiliation{\atOSU}
\altaffiliation{\atChicago}

\author{C.~Miki}
\affiliation{\atUH}

\author{J.~Nam}
\affiliation{\atNTU}

\author{C.~J.~Naudet}
\affiliation{\atJPL}

\author{R.~J.~Nichol}
\affiliation{\atUCL}

\author{A.~Novikov}
\affiliation{\atKU}
\affiliation{National Research Nuclear Univ., Moscow Engineering Physics Inst., Moscow, Russia.}

\author{E.~Oberla}
\affiliation{\atChicago}

\author{M.~Olmedo}
\affiliation{\atUH}

\author{R.~Prechelt}
\affiliation{\atUH}

\author{B.~F.~Rauch}
\affiliation{\atWashU}

\author{J.~M.~Roberts}
\affiliation{\atUH}

\author{A.~Romero-Wolf}
\affiliation{\atJPL}

\author{B. Rotter}
\affiliation{\atUH}

\author{J.~W.~Russell}
\affiliation{\atUH}

\author{D.~Saltzberg}
\affiliation{\atUCLA}

\author{D.~Seckel}
\affiliation{\atUD}

\author{H.~Schoorlemmer}
\affiliation{\atMPI}

\author{J.~Shiao}
\affiliation{\atNTU}

\author{S.~Stafford}
\affiliation{\atOSU}

\author{J.~Stockham}
\affiliation{\atKU}

\author{M.~Stockham}
\affiliation{\atKU}

\author{B.~Strutt}
\affiliation{\atUCLA}

\author{M.~S.~Sutherland}
\affiliation{\atChicago}

\author{G.~S.~Varner}
\affiliation{\atUH}

\author{A.~G.~Vieregg}
\affiliation{\atChicago}

\author{S.~H.~Wang}
\affiliation{\atNTU}

\author{S.~A.~Wissel}
\affiliation{\atPennState}

\vspace{2mm}
\noindent

\begin{abstract}
ANITA's fourth long-duration balloon (LDB) 
flight in 2016 detected 29 cosmic-ray (CR)-like events on a background of 
$0.37^{+0.27}_{-0.17}$ anthropogenic events.
CRs are mainly seen in reflection off the Antarctic ice sheets, creating
a phase-inverted waveform polarity. However, four of the below-horizon CR-like events show anomalous non-inverted polarity,
a $p = 5.3 \times 10^{-4}$ chance if due to background. All anomalous events are from locations near the horizon;
ANITA-IV observed no steeply-upcoming anomalous events similar to the two such events seen in prior flights.
\end{abstract}

\maketitle

Antarctic ice has been recognized for decades as an ideal natural dielectric target for the detection of 
cosmic neutrinos via the
Askaryan effect, which leads to coherent radio Cherenkov impulses from particle cascades in
dielectric materials~\cite{Ask62,SLAC01,T486}. The ANtarctic Impulsive Transient Antenna (ANITA) instrument
was designed to exploit this effect by broadband monitoring of several million
km$^3$ of ice in the 200-1200~MHz band from the stratosphere during a LDB flight~\cite{ANITA-inst}. 

During prior flights, 
ANITA also found that several dozen ultra-high energy (UHE) cosmic ray (CR) events were also detectable from the
payload's stratospheric vantage point~\cite{ANITA-CR-2010}. 
Cosmic ray extensive air showers (EAS) in the geomagnetic field produce 
$1^{\circ}$-beamed radio impulses
via a charge acceleration mechanism tied to the magnetic Lorentz force $\mathbf{F}_{\mathcal{L}} = q \mathbf{v} \times \mathbf{B}_{geo}$
for particle charge $q$, velocity $\mathbf{v}$, and geomagnetic field $\mathbf{B}_{geo}$.
In CR EAS this mechanism dominates over the 
Askaryan effect, leading to signals with strong correlations in polarization to the local geomagnetic field 
where the shower propagated.

For the fourth flight of ANITA, we
pursued two separate analysis paths, one for neutrinos interacting below the ice surface
and detected via the Askaryan channel, and one for CR or
CR-like events, detected via the EAS channel. 
Results of the ANITA-IV analysis for neutrino events via the Askaryan channel
are reported elsewhere~\cite{A4neutrino}.

\begin{figure}[htb!]
\vspace{-2mm}
\begin{center}
\centerline{\includegraphics[width=3.3in]{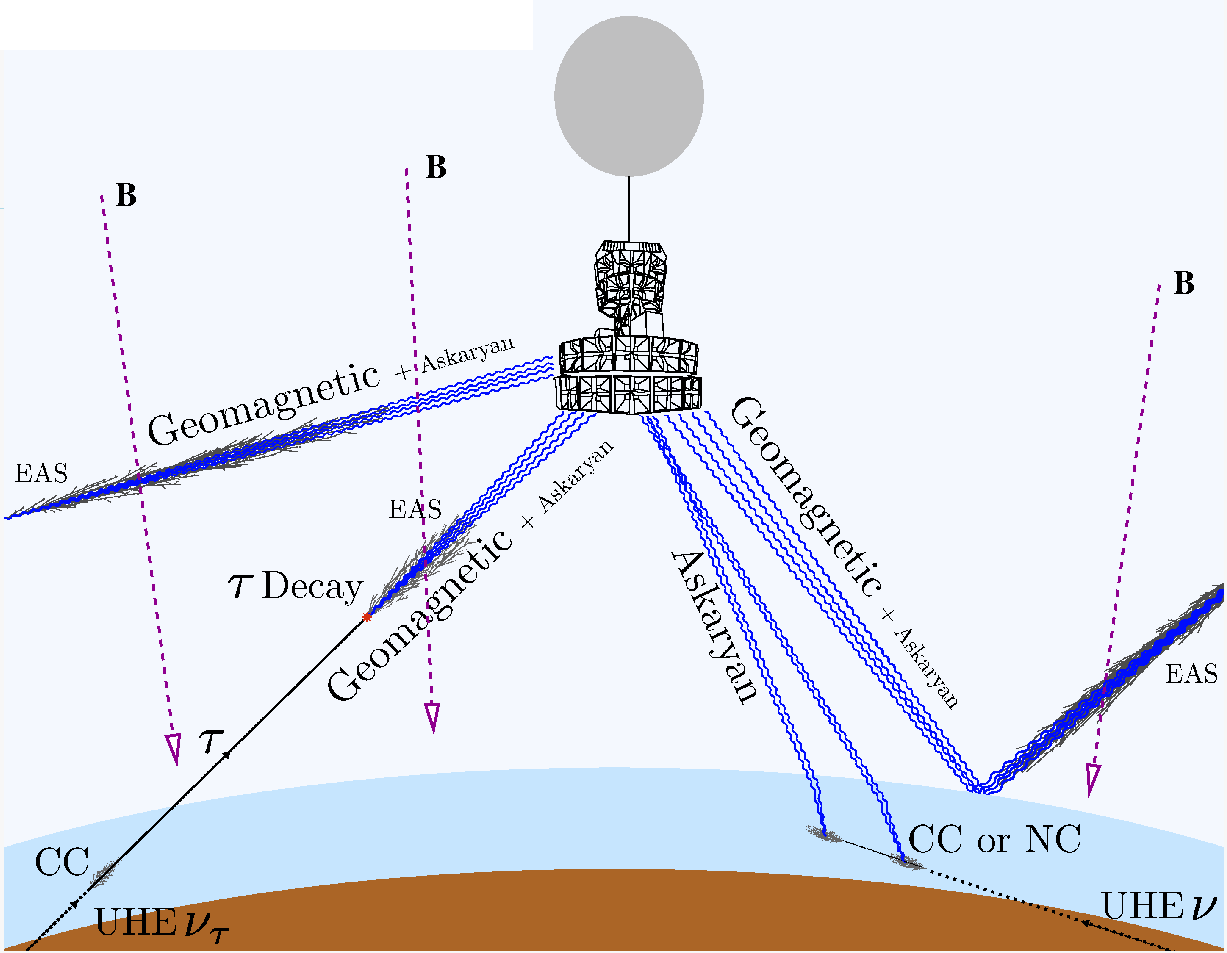}}
\caption{\small \it Diagram of ANITA-IV signal channels, in near-vertical
Antarctic geomagnetic field. Geomagnetic signals
appear from CR either directly or in reflection, and include a small Askaryan
radiation component. Askaryan signals dominate the under-ice neutrino channel,
and can arise from multiple showers due to both primary and secondary
interactions. Geomagnetic signals can also arise from EAS initiated by $\tau$-lepton-decay 
after $\nu_{\tau}$ charged-current (CC) interactions within the ice.
\label{A4diagram}}
\vspace{-9mm}
\end{center}
\end{figure}

ANITA's effective area for CR
detection is not competitive with other large ground-based CR observatories~\cite{Harm16}, 
but its field-of-view from the 
stratosphere does provide access to geometries that ground-based CR observatories cannot see.
CRs that arrive tangential to Earth's surface may interact in the stratosphere, and
the EAS may be confined completely to the stratosphere,
never intersecting with Earth's surface. Such stratospheric events appear to ANITA as a CR
radio source near, but just above the horizon, in a thin band of atmosphere observed
at the limb of the Earth. In contrast, the large majority of CRs which arrive on trajectories that
do intersect the Earth appear to ANITA in reflection off the relatively radio-smooth surface of the ice.
The reflection causes a phase inversion in the CR waveform, providing a clear distinction between
these down-going CRs and their stratospheric counterparts. Figure~\ref{A4diagram} provides a schematic
overview of the various possible signal channels for ANITA.

In each of two prior ANITA flights where CR observations were made, single anomalous CR
events were observed at payload arrival angles of $-27^{\circ}$~\cite{Upshowers} and  $-35^{\circ}$~\cite{A3upwardshower}
relative to horizontal.  The polarity of these events was not phase-inverted as 
was the case for all of the other
(several dozen) reflected CRs observed in the below-horizon angular region. 
These CR-like events had low probability of being background,
especially for the unusual event observed by ANITA-III~\cite{A3upwardshower}.
No natural origin for the non-inverted polarity has yet been
confirmed, though several have been suggested~\cite{Stoprad,Glacial}, and in some cases,
largely excluded by ANITA~\cite{Hical20}. 
IceCube has also published constraints on potential astrophysics
sources for these events~\cite{IC2020}.

ANITA-IV was launched on Dec.~2, 2016, reaching a float altitude of
about 40~km several hours later, and flew in the Antarctic polar vortex for 28~days until 
the flight was terminated on Dec.~29, 2016, about 160~km from the South Pole.
The data recording livetime averaged $\sim 90$\% during the flight, yielding 24.5~days net.

The blind analysis~\cite{LudwigThesis} used to extract the ANITA-IV CR sample 
followed closely the methods detailed for ANITA-III~\cite{A3upwardshower}, and
those described for the neutrino analysis~\cite{A4neutrino}. The CR analysis methods included using
cross-correlation with a CR waveform template and polarization correlation with the local geomagnetic field.

\begin{figure}[htb!]
\vspace{-2mm}
\begin{center}
\centerline{\includegraphics[width=3.5in]{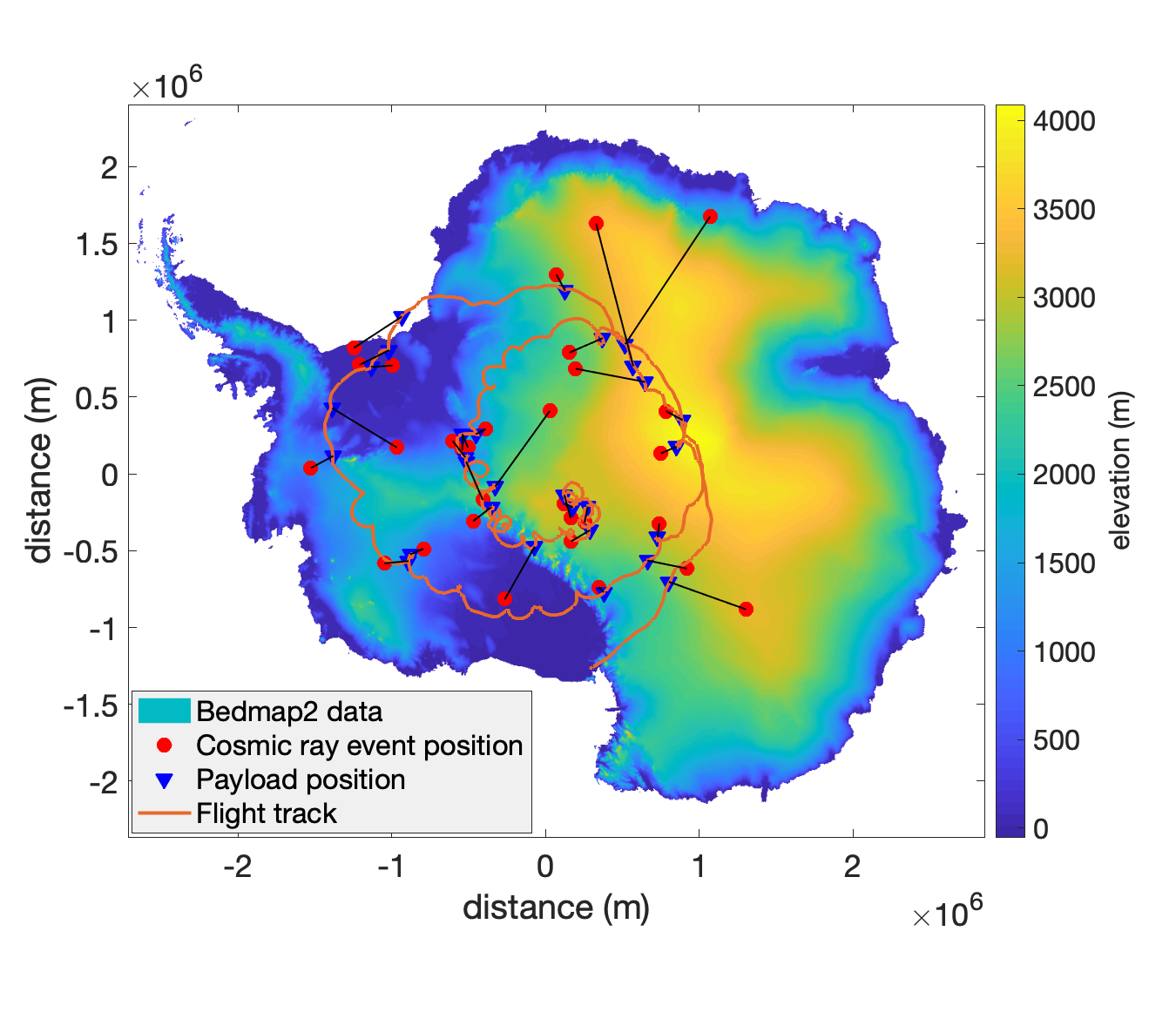}} 
\centerline{\includegraphics[width=3.05in]{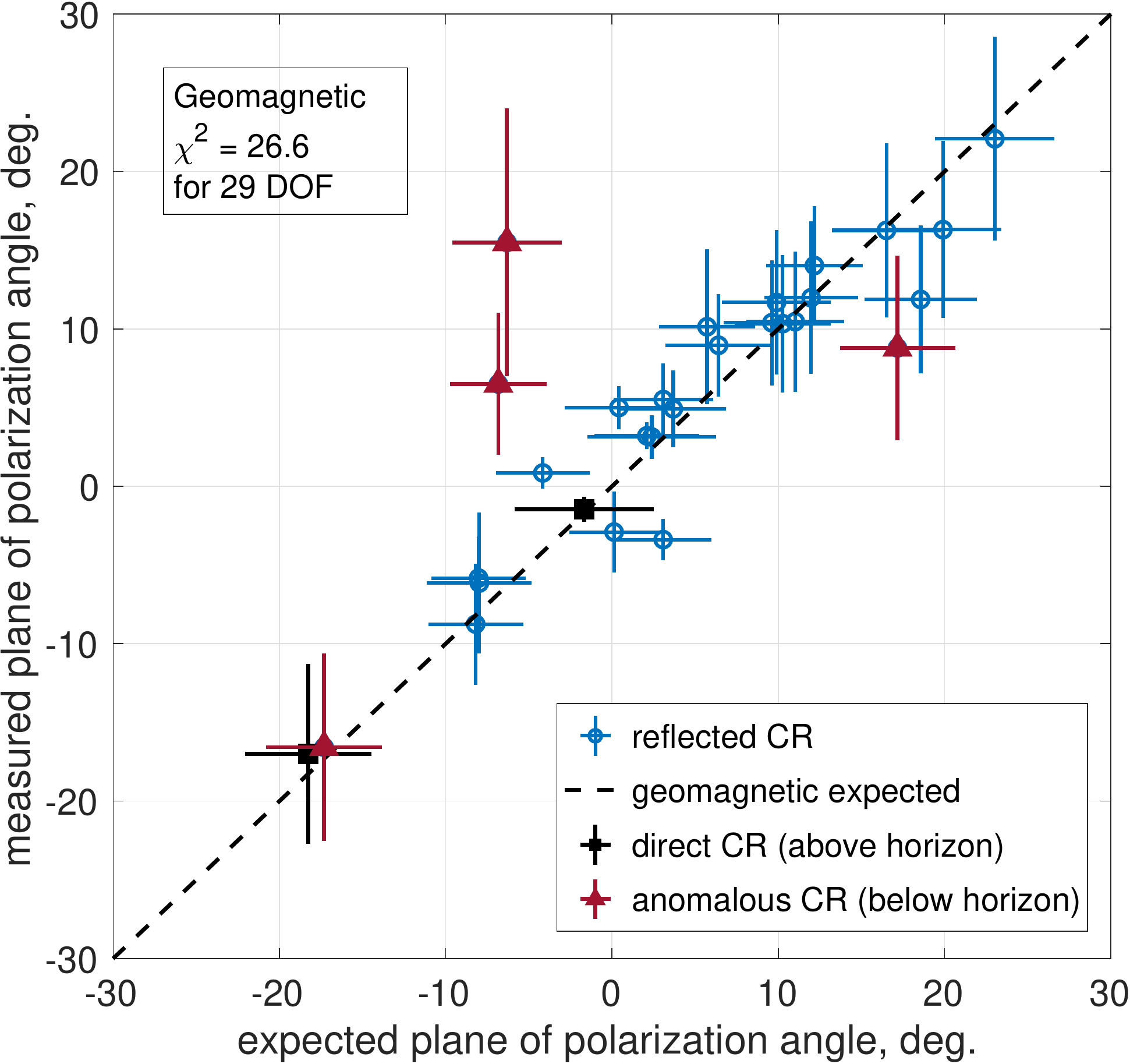}}
\caption{\small \it Top: ANITA-IV flight path and location of payload and
apparent event source location and ice surface elevation above sea level for each of the 29 events in the final CR sample.
Bottom: Geomagnetic correlation of 29 candidate events revealed in our
CR analysis.
\label{A4map}}
\vspace{-9mm}
\end{center}
\end{figure}

%
Events from large active bases: McMurdo, WAIS, South Pole, etc.
are excluded as signal.
A statistical sideband sample, extracted from excluded events with known anthropogenic
associations, was used to estimate the anthropogenic background,
which is assumed to ``leak'' out from weak sources into isolated single events. 
The final signal analysis is still blind to
polarity at this stage, and a detailed analysis of the anthropogenic sample gives a background estimate 
of $0.37^{+0.27}_{-0.17}$ events, for events of both polarities in the final
CR sample. By comparison the thermal radio
noise background contribution is negligible, $\sim 5 \times 10^{-7}$~events.

The complete unblinded sample of isolated events showed 29 candidates
distributed widely across the continent, as shown in Fig.~\ref{A4map}~(top)~\cite{AntarcticMap,Bedmap2}.
They are consistent with CRs in their template correlation coefficient and geomagnetic parameters (Fig.~\ref{A4map}, bottom).
The events were observed
at arrival angles of $-36^{\circ}$ to $-5.5^{\circ}$ with respect to horizontal (the horizon
appears at about $-6^{\circ}$ relative to horizontal from stratospheric altitudes). The two events observed
from above the horizon are identified as candidate stratospheric EAS, a class of extensive air 
showers first observed by ANITA~\cite{ANITA-CR-2010}.

Prior to determining polarity, the events are processed to form 
coherently-summed waveforms (CSW), created by combining signal from
15 antennas that contain the source within their field-of-regard,
with group delays that match the source direction.  This process
is known as {\it beamforming} in radio interferometric usage. 
ANITA's beamforming typically improves the signal-to-noise ratio (SNR) by a factor of 
three compared to the original detection.

ANITA's CSW show a time-dispersed shape caused by the variable
group delay of different frequency components passing through the ANITA 
antenna, receiver, and digitizer systems. Each intrinsic CR impulse is therefore convolved with
this system impulse response, inducing both phase and amplitude distortion
in the received signal. This impulse response also varies during the flight,
due to changes in the frequency-notch configuration as we work to suppress
electromagnetic interference~\cite{PaperI}.

\begin{figure}[htb!]
\vspace{-3mm}
\begin{center}
\centerline{\includegraphics[width=3.85in]{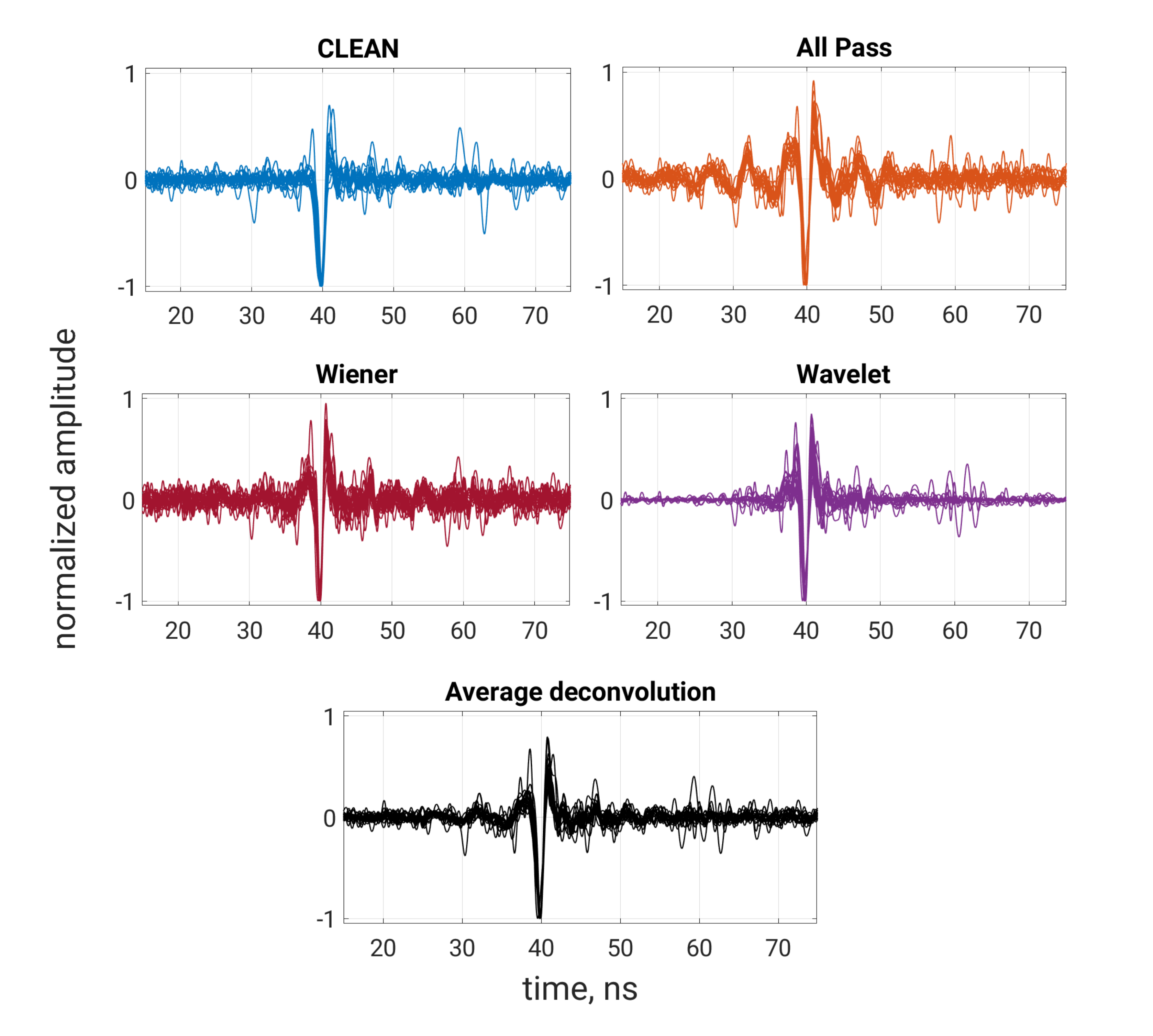}}
\vspace{-3mm}
\caption{\small \it Normalized overlays of the 21 normal reflected CR in our sample,
in each of the four deconvolution methods used, as noted by the title for each pane.
CLEAN~\cite{CLEAN1} is a time-domain method based on iterative correlation and subtraction of the 
impulse response~\cite{CLEAN2,CLEAN3}. The All-pass method applies 
simple Fourier-division deconvolution to the phases only.
Wiener deconvolution uses a noise-optimized Fourier deconvolution method 
for both phases and amplitudes~\cite{Wiener}. The Wavelet method~\cite{ForWaRD04, Debdeep} uses
wavelet, rather than Fourier, basis functions for the deconvolution.
The bottom plot gives the normalized average waveform of the four methods.}
\label{allCRoverlay}
\end{center}
\vspace{-6mm}
\end{figure}

Calibration errors in portions of the global analysis discovered after the
initial CR unblinding required that the data be reanalyzed~\cite{PaperI}.
The final subsequent polarity analysis was done after reblinding of the data, including blinding of 
event number, randomizing event order, and applying a random polarity factor to 
all CR events as the polarity metrics were determined.

Polarity is unrelated to the polarization state of the event.
Polarity refers to the instantaneous phase of the electric field, 
whereas polarization refers to the plane of the field oscillation. 
After deconvolution of the system response~\cite{PaperI}, CR events observed
by ANITA vary from nearly unipolar, to dominant bipolar, and some subdominant tripolar events.

Polarity of a unipolar pulse is determined
by the sign of the pole. In bipolar events, the polarity is indicated by the order of the two primary poles;
for ANITA data, we find that the sign of leading main pole in a bipolar event determines
the polarity. Tripolar components for ANITA events are always subdominant to unipolar or bipolar
shapes, and do not affect the polarity determination in general.

After unblinding polarity we find two events with indeterminate polarity; 
coincidentally, both
had payload arrival angles of  $-14.8^{\circ}$. After further investigation,
one of these events, 74197411, was found to be adversely affected by
high-frequency interference of likely
anthropogenic origin. Filtering this interference revealed that the
polarity was that of a normal CR. 
For the
other event, 88992443, the polarity could not be resolved, and remains
indeterminate. This event has a waveform quite
different from all other CR, and has the lowest SNR of any CR
observed. Both of these events are excluded from these analyses;
their data is provided elsewhere~\cite{PaperI}.

For the remaining 27 events,the confidence level (CL) for polarity determination was found
by Monte Carlo methods to be $\sim 99\%$ for one event (19848917),
and $\geq 99.99\%$ for the remainder. For 19848917, the $\sim 1\%$ 
chance of polarity misidentification appears to be due to limitations of our
algorithm rather than intrinsic uncertainty in this event's polarity.

Of the remaining 27 events, 23 events have the normal CR polarity expected from their geometry.
Figure~\ref{allCRoverlay} shows an overlay of the 21 events with
polarity consistent with reflection from the ice sheet surface, the
most common type of CR observed by ANITA. Each of the panes shows
the results of the different deconvolution methods~\cite{PaperI}
along with an overlay of the average of all four waveform methods.
This final waveform average, chosen for robustness to systematics in the individual
deconvolutions, was used to determine the polarity in all events.

\begin{table*}[htb!]
\begin{threeparttable}
\caption{\footnotesize Preliminary list of stratospheric CR and 
possible anomalous CR-like events seen by ANITA-IV.}
\label{MEtable}
\centering
\begin{footnotesize}
 \begin{tabular}{|c|c|c|c|c|c|c|c|c|}
 \hline
 
 event \# & mm dd hh mm ss & {Apparent source location} & elev. angle\tnote{a}& horizon angle\tnote{a~}~& azimuth &{Payload location} &   Type\tnote{c~}~ & Energy\tnote{d~}  \\
        &  UTC 2016  & Lat.$^{\circ}$,Lon.$^{\circ}$, alt., m & degrees &  degrees & degrees  &  Lat.$^{\circ}$, Lon.$^{\circ}$, alt., km &   & EeV \\ \hline
4098827 &  12 03 10 03 27 & -75.71, 123.99, 3184  &  $-6.17\pm 0.21$ &   $-5.92\pm 0.020$   & 337.70 & -80.157, 131.210,  38.86 & NI &  $1.5 \pm 0.7$   \\  
9734523  & 12 05 12 55 40 & -71.862, 32.61, 19000\tnote{b}  &  $-5.64\pm 0.20 $  &   $-5.95\pm 0.020$ &  2.01 & -80.9, 31.6, 39.25 &  AH &  ...   \\
19848917 & 12 08 11 44 54 & -80.818 , -79.87, 758 & $-6.71 \pm 0.20$ & $-6.06 \pm 0.020$ & 194.34 & -76.66, -72.86,  38.97  &   NI  &  $0.9 \pm 0.5$    \\
50549772 & 12 16 15 03 19 & -83.483, 14.73, 2572  & $-6.73 \pm 0.20$ & $-5.92\pm 0.020$  & 234.08 & -81.95, 47.29, 38.52 &   NI  &  $0.8 \pm 0.3$   \\
51293223 & 12 16 19 08 08 & -74.800, 11.43, 18600\tnote{b} &  $-5.38\pm 0.24$  &   $-5.85\pm 0.020$ & 306.45 &  -81.7, 39.2, 37.53 &  AH  & ...\\
72164985 & 12 22 06 28 14 & -86.598, 0.35, 2589 &   $-6.12\pm 0.10$ & $-5.93\pm 0.020$  & 140.03 & -86.93, -104.29,  38.58 &  NI &   $3.9 \pm 2.5$ \\ 
\hline
 \end{tabular}
 \end{footnotesize}
 \begin{tablenotes}
  \footnotesize
  \item[a]  Both the observed elevation angle and the apparent horizon here include radio refraction, which lifts the apparent 
  horizon about 0.1$^{\circ}$. 
  \item[b] The source elevation (in the stratosphere) and the given source position are estimates of the approximate location of EAS maximum for
  these direct stratospheric CR events, determined by using the average column depth to shower max for EeV CRs.
  \item[c] AH: above-horizon, direct CR. NI: Non-inverted CR-like event, below horizon.
  \item[d] Energy computable for below-horizon events only; above-horizon simulations are beyond our scope in this report. Errors include
    both statistical and systematic effects.
 \end{tablenotes}
\end{threeparttable}
\vspace{-6mm}
\end{table*}

In addition to the 23 normal reflected CR, four events near the horizon 
had non-inverted polarity, inconsistent
with a reflected CR, though their source directions were on the ice sheets.
Table~\ref{MEtable} shows parameters for the four near-horizon and two above-horizon events. 
Pointing parameters in these cases used
weighted averages of the pointing determined from interferometric maps of both polarizations where
there was sufficient SNR. Under the assumption these are CR showers, the table includes estimates of the energy based on scaling from our prior CR energy measurements~\cite{Harm16}. 

The stratospheric events 9734523 and 51293223 are consistent with CRs 
developing at altitudes of 18-19~km above
the Earth's surface. As such they enter the Earth's atmosphere at distances beyond the 
physical horizon, and the resulting particle cascades develop over hundreds of km through the rarefied
stratosphere at such altitudes. 
Estimation of their energies will have to await detailed simulations of these
rather extreme form of EAS.
These two above-horizon
events appear at angles of $0.3^{\circ}$ and $0.47^{\circ}$ above the horizon,
respectively. In ANITA-III we observed one other stratospheric event at $\sim 0.4^{\circ}$
above the horizon, and several at larger angles, but none closer than this.

\begin{figure}[htb!]
\begin{center}
\vspace{3mm}
\centerline{\includegraphics[width=3.45in]{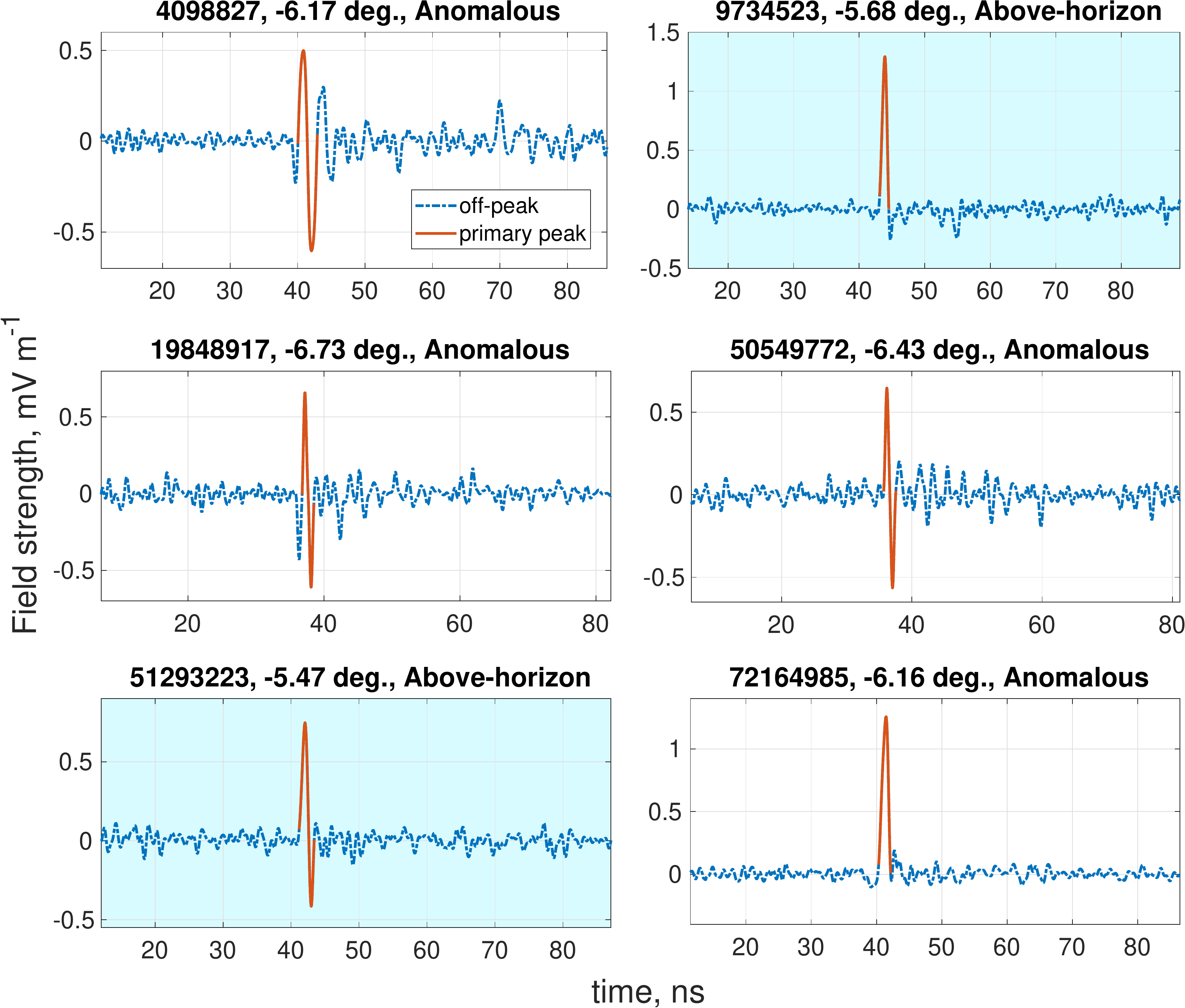}}
\caption{\small \it The incident field strength vs. time for the six near-horizon events,
all of which have the same (non-inverted) polarity:
two above-horizon (pale blue background), and four below-horizon
(white background) with anomalous polarity. 
These plots use the CLEAN deconvolution method as the waveform estimate.}
\label{sixevents}
\end{center}
\vspace{-10mm}
\end{figure}

For the events summarized in Table~\ref{MEtable},  
Fig.~\ref{sixevents} shows the CLEAN deconvolutions 
scaled to give their incident electric field strength at the ANITA payload. 
Here the subdominant Vpol waveform
component was added coherently to the Hpol component to produce the best estimate
of the intrinsic field strength of the incident pulse in the plane of polarization.
Systematic errors on the field strength are estimated to be $\sim 30$\%. 
The plot background colors indicate events from above horizon (pale blue) or below horizon (white).
The central segments of the waveforms
annotated in solid orange show the primary pole (for unipolar) or poles 
(for bipolar) that determine the polarity; dot-dash lines show the off-peak waveforms and thermal noise.
In laboratory tests using known signals plus thermal noise, CLEAN gave the most
consistent recovery of the intrinsic waveform in the presence of noise, and thus we use
the CLEAN waveform for the estimates in Fig.~\ref{sixevents}.

Event 4098827 had significant loss of bandwidth due to the three notch filters used to suppress anthropogenic
interference for this event, and although the deconvolution recovers a portion of this signal via interpolation,
the narrower effective bandwidth of this event is reflected in the 
larger number of zero-crossings in the resulting waveform.
Despite this, 4098827's polarity is in this case known with confidence, 
due to fortuitous observation of a 
normal three-notch-filter CR event~\cite{PaperI} with a very similar waveform,
and which had inverted polarity with respect to this event.

In addition to the dedicated CR analysis, a separate neutrino-focused analysis chain based only
on the impulsive nature of events and spatial isolation of their locations also found the majority of the CRs, independently confirming the efficacy of the CR methods~\cite{PengThesis}; further details
are given in supplementary material~\cite{PaperI}.



In Table~\ref{MEtable}, events 4098827 and 72164985 
just below the horizon, by about $0.2^{\circ}$, about 1 and 2 standard deviations,
respectively, for these two events.
The standard error in angle given depends on their SNR, determined 
from ground-to-payload calibration pulsers during the flight.
Both of these events thus have a non-negligible statistical chance to be due to
stratospheric CRs  misidentified as below-horizon events.
To assess the chance of such a misidentification, we must 
consider the effects of grazing incidence RF propagation in the 
near-horizon atmosphere.

A detailed analysis of the effects of near-horizon propagation, including the use of GPS
occultation data~\cite{GPSocc, gps2}, indicates that the refractivity gradient 
can lead to significant phase distortion if the ray paths fall within 1~km of the surface,
a region that is comparable to the size of the first Fresnel zone
for ANITA's geometry~\cite{PaperI}. Although we do not have direct measurements of the tropospheric parameters
in the region that these events were observed, measurements taken at the South Pole indicate 
that spatial variations in the index of refraction observed in this near-surface atmosphere
would lead to variations in the path delay across the wavefronts and loss of coherence at higher frequencies.

Such a loss of coherence does not appear consistent 
with our observations~\cite{PaperI}. This suggests that if 4098827 and 72164985
arise from misidentified above-horizon CRs, they require a true source direction that is
$\gtrsim 0.1^{\circ}$ above the horizon to preserve the observed coherence of the events. 
This requirement reduces the likelihood of an origin due to pointing fluctuations. 

In addition to these propagation considerations, we have also considered the effects of a residual unknown
$\pm 0.1^{\circ}$ systematic pointing bias that might offset the apparent location of the events
relative to their true direction. Evidence from our calibration and other 
data does not exclude possible bias at this level.

We exclude one effect which can produce grazing-incidence, below-horizon
images of an above-horizon source: the so-called {\it inferior mirage} effect. This well-known
manifestation is due to a near-surface refractive inversion layer. 
Under typical conditions it produces inverted
reflections, inconsistent with our observations. There
are however reports of non-inverted inferior mirages; careful study of the optics of
these~\cite{vanderWerf2011} show that such images are the result
of coherent surface concavities on a scale very long compared to the wavelength.
In effect, the surface curvature mimics a concave ellipsoidal mirror, with one 
of its foci between the source and observer, thus causing an additional image inversion
that cancels the reflection-inversion. We have carefully searched ice-surface altimetry
data~\cite{Bedmap2} for evidence of such features for each of the four events, and none show surface
altimetry consistent with this.

To estimate the overall significance of observing these four events given 
the anthropogenic background estimate, the chance for 
statistical polarity misidentification, and the chance for misreconstruction of the event direction,
we performed two independent statistical simulations, 
each of which vary parameters for all 27 CR with determined polarities, tabulating how
often the outcomes randomly produce four or more such events in any combination. We allow variations for all
types of systematic error noted above: a pointing bias,
a propagation restriction, and polarity misidentification, 
to provide conservative bounds on the significance.

Both simulations gave consistent results, indicating a $p$-value range of 
$$(3.7 \times 10^{-3}) \geq p \geq (7.5 \times 10^{-5})$$
equivalent to $(3.3 \pm 0.5)\sigma$ significance in Gaussian statistics~\cite{PaperI}. 
While this CL is
not adequate to conclude that these events may not be some combination of the different backgrounds,
it is suggestive of a new class of CR-like events with Earth-skimming geometry.

ANITA-IV's sensitivity exceeded 
that of any prior ANITA flight. Three of these four events are near the threshold of sensitivity,
and would not have been observed previously. Currently, there is no
radio EAS simulation that can treat events propagating in these very extreme conditions where
a full-wave physical-optics solution is necessary. Such full-wave simulations
with scales that can match the ANITA geometry are currently beyond the computing capabilities 
available for this work, and will require a follow-up investigation to understand ANITA's acceptance to
near-horizon EAS, and any relevant near-horizon propagation effects that may impact the significance
of the observation. 

If such events arise from unexpected, but potential mundane physical optics in the 
cryosphere, or from other atmospheric or ice-topographic effects, it is crucial for these effects,
which constitute a potential background for future experiments, be explored in detail. 
Orbital and suborbital missions such as POEMMA~\cite{POEMMA} and PUEO~\cite{PUEO},  and ground-based 
experiments such as GRAND~\cite{GRAND} which will specifically explore 
the potential $\tau$-lepton neutrino channel with much higher sensitivity, 
stand to benefit from a better understanding of all aspects of this process.

While the significance of the data does not yet require it, 
we anticipate the question of a possible particle physics origin
for these events. The four near-horizon events are not inconsistent with 
EAS initiated by $\tau$-lepton decay after emergence of the $\tau$ from a charged-current neutrino event
in the ice along the track direction. Because their track directions are near tangential,
the parent $\nu_{\tau}$ would not suffer significant attenuation in the Earth, a problem that appeared to exclude
a Standard-Model neutrino origin for the steeply-arriving anomalous 
events observed in earlier ANITA flights~\cite{Taupaper}.  

Thus it is possible that one or more of these events can arise from an
EAS generated by $\tau$-lepton decay. However, to avoid saturating diffuse flux bounds,
a transient point-source neutrino fluence is more likely than a steady-state diffuse flux,
as is commonly assumed for UHE cosmogenic neutrinos.
ANITA's point-source effective area for $\tau$-lepton
generated EAS is maximal at the horizon, of order $0.1$~km$^2$ at EeV energies, including
the neutrino cross section and $\tau$ survival and decay effects. The implied UHE 
integral $\nu_{\tau}$ fluence in this case, assuming
an astrophysical transient event with a timescale of 1000~s or less, is of order 10
per km$^2$ above 3 EeV, which appears allowed by all current bounds on transients at these energies. 
Further work on understanding potential sensitivity to such
events by other experiments is forthcoming.


\begin{acknowledgments}
ANITA-IV was supported by NASA grant NNX15AC24G and related grants.  We thank the staff of the 
Columbia Scientific Balloon Facility for their generous support.  
This work was supported by the Kavli Institute for Cosmological Physics at the University of Chicago.
Computing resources were provided by the Research Computing Center at the University of Chicago and the 
Ohio Supercomputing Center at The Ohio State University.
O. Banerjee and L. Cremonesi's work was supported by collaborative visits funded by the Cosmology and 
Astroparticle Student and Postdoc Exchange Network (CASPEN).
S. A. Wissel would like to thank the Cal Poly Frost Fund and the CSU Research, Scholarship, and Creative Activity (RSCA) Grant Program.
The University College London group was also supported by the Leverhulme Trust. 
The National Taiwan University group is supported by Taiwan's Ministry 
of Science and Technology (MOST) under its Vanguard Program 106-2119-M-002-011.
\end{acknowledgments}

\end{document}